\documentclass[twocolumn, tighten, times, trackchanges]{aastex631}

\usepackage{amsmath}

\usepackage{hyperref}
\usepackage{multirow}

\shorttitle{Exploring the Emission Mechanisms of Mrk 180}
\shortauthors{Mondal et al.}

\begin{document}

\vspace*{-1.25cm}

\title{Exploring the Emission Mechanisms of Mrk 180 with long term X-ray and $\gamma$-ray data}

\author[0000-0003-2445-9935]{Sandeep Kumar Mondal}
\affiliation{Astronomy \& Astrophysics Group, Raman Research Institute, C.V. Raman Avenue, Sadashivanagar, Bangalore 560080,  Karnataka, India}
 \email{skmondal@rri.res.in}

\author[0000-0001-5796-225X]{Saikat Das}
\affiliation{Yukawa Institute for Theoretical Physics, Kyoto University, Kitashirakawa Oiwakecho, Sakyo-ku, Kyoto 606-8502, Japan}
\email{saikat.das@yukawa.kyoto-u.ac.jp}

\author[0000-0002-1188-7503]{Nayantara Gupta}
\affiliation{Astronomy \& Astrophysics Group, Raman Research Institute, C.V. Raman Avenue, Sadashivanagar, Bangalore 560080,  Karnataka, India}
\email{nayan@rri.res.in}


\begin{abstract}
Markarian (Mrk) 180 is a BL Lacertae (BL Lac) object located at a redshift of 0.045 and a potential candidate for high-energy cosmic ray acceleration. We have analyzed the Fermi Large Area Telescope (\textit{Fermi}-LAT) $\gamma$-ray data of Mrk 180 collected over a period of 12.8 years and found no significant enhancement in the flux from the long-term $\gamma$-ray light curve. We have also analyzed Swift X-ray, ultraviolet \& optical, and X-ray Multi-Mirror Mission (XMM-Newton) data to construct the multi-wavelength spectral energy distribution (SED). The SED has been modeled with one-zone pure leptonic and lepto-hadronic scenarios to explain the underlying physics of multi-wavelength emission. The pure leptonic model and the two lepto-hadronic models, viz., (i) line-of-sight interactions of ultrahigh-energy cosmic rays (UHECR; $E\gtrsim 10^{17}$ eV) with the cosmic background radiation and (ii) the interactions of relativistic protons with the cold protons in the jet, have been compared in our work. Moreover, an earlier study has associated Mrk 180 with the Telescope Array (TA) hotspot of UHECRs at $E>57$ EeV. This speculation motivates us to check whether ultrahigh energy protons and iron nuclei can reach the earth from Mrk 180. After comparing the results of our simulation with the current observational data, we find that Mrk 180 is unlikely to be a source of the UHECR events contributing to the TA hotspot for conservative strengths of extragalactic magnetic fields. 
\end{abstract}


\section{Introduction} \label{sec:intro}
The central emission core of active galaxies is powered by accretion onto a supermassive black hole (SMBH). This leads to the formation of a collimated jet of outflow, along the angular momentum direction, that outshines the entire galaxy \citep{Urry_1995}. Active Galactic Nuclei (AGNs) are one of the most prominent sources of high energy $\gamma$-rays. The jet transports energy and momentum over large distances. In the case of blazars, the jet points along the observer's line of sight and provides a unique testbed to study the acceleration of cosmic rays \citep[see][for a recent review]{Blandford:2018iot}. Blazars show high flux variability; their emission is highly polarised and the emission is of non-thermal origin. They are broadly classified into flat spectrum radio quasars (FSRQs) having broad emission lines, and BL Lac objects showing a featureless continuum spectrum.

The broadband SED of a blazar covers the entire electromagnetic spectrum, ranging from radio to very high-energy (VHE, $E\gtrsim 30$ GeV) $\gamma$-rays. It exhibits two peak emission frequencies. The low-energy peak occurs between radio to soft X-ray energies and can be attributed to synchrotron radiation of relativistic electron and positron population. The high-energy peak between X-ray to VHE $\gamma$-ray energies can arise from various  processes. The most prevalent explanation is the inverse-Compton scattering of synchrotron photons (synchrotron self-Compton, SSC) or external photons (EC) originating from the broad-line region (BLR), dusty torus (DT), or the accretion disk (AD). In addition, the VHE $\gamma$-rays can also come from photohadronic ($p\gamma$) or hadronuclear ($pp$) interactions of accelerated cosmic rays with the ambient radiation or matter in the emission region of the jet or proton synchrotron radiation \citep{2014NIMPA.742..191T, Blandford:2018iot, Cerruti:2020lfj}.

Mrk 180 was discovered by Swiss-origin astronomer Fritz Zwicky and later identified as a BL Lac object in 1976 by spectral analysis. It is a high-synchrotron peaked BL Lac (HBL) object embedded at the center of an elliptical galaxy \citep{1981ApJ...248L..61M}, located at redshift, z=0.0458 \citep{1978ApJ...222L...3U} with R.A.= 174.11008 deg, Decl.= 70.1575 deg. This source was detected for the first time in X-rays by HEAO-1 \citep{1981AJ.....86.1585H}, since then it has been monitored by several telescopes e.g. \textit{Fermi}-LAT, Swift, Major Atmospheric Gamma Imaging Cherenkov Telescope (MAGIC), XMM-Newton, Monitoring of jets in Active Galactic Nuclei with VLBA Experiments (MOJAVE), KVA, ASM. In March 2006, VHE $\gamma$-ray emission was detected for the first time \citep{Albert_2006} from this source, triggered by an optical burst. \cite{https://doi.org/10.48550/arxiv.1110.6341} and \cite{https://doi.org/10.48550/arxiv.1109.6808} did multi-wavelength study on this source. Mrk 180 was also monitored for a long period (2002- 2012) in the optical waveband and its light curve was analyzed \citep{Nilsson_2018}.

The Telescope Array experiment, located in Utah, United States, is a state-of-the-art detector observing ultrahigh-energy cosmic rays (UHECRs; $E\gtrsim 10^{17}$ eV) in the northern hemisphere. Based on an intermediate-scale anisotropy search using 5 years of data, the TA collaboration had earlier reported a cluster of events at RA=146$^\circ$.7 and Dec=43$^\circ$.2, found by oversampling in 20$^\circ$ radius circles \citep{TelescopeArray:2014tsd}. 72 UHECR events were detected in this direction at $E>57$ EeV, where TA has 100\% detection efficiency. The hotspot had a Li-Ma significance of 5.1$\sigma$. \cite{PhysRevD.93.043011} identified Mrk 180 as a possible source of UHECRs in the context of explaining the origin of the TA hotspot \citep{TelescopeArray:2014tsd, 2015ICRC...34..276K, 2019ICRC...36..310K}. Motivated by the earlier studies, we carry out a comprehensive study of Mrk 180 to ascertain the underlying mechanism of high-energy $\gamma$-ray emission and whether it can be the source of UHECRs beyond 57 EeV contributing to the TA hotspot.

We have analyzed the \textit{Fermi}-LAT data collected over a period of 12.8 years, the Swift XRT and UVOT data, and in addition to these the XMM-Newton X-ray data to construct the broadband SED of this source. Section~\ref{sec:data} is dedicated to the discussions on the methods followed to analyze the data. We have also searched for fluctuations in the $\gamma$-ray flux in the \textit{Fermi}-LAT light curve, as discussed in Sec.~\ref{subsec:light_curve}. Subsequently, we build the long-term multi-wavelength SED. We discuss the theoretical framework for SED modeling in Sec.~\ref{sec:sed_model}. We present our results in Sec.~\ref{sec:results} and discuss them in Sec.~\ref{sec:discussion}. Finally, we draw our conclusions in Sec.~\ref{sec:conclusion}.


\section{ \label{sec:data}Data Analysis}


\subsection{\label{subsec:fermi}Fermi-LAT Data Analysis}

The \textit{Fermi}-LAT is an imaging, pair-conversion, wide-field-of-view, high-energy $\gamma$-ray telescope that can detect photons of energy 20 MeV to more than 300 GeV, whose field of view is 2.4 sr \citep{Atwood_2009}. Fermi carries two instruments, one is the LAT and the other one is the Gamma-ray Burst Monitor (GBM). The LAT is Fermi's primary instrument. Fermi scans the whole sky every three hours. It was launched in June 2008 in the near-earth orbit and still in operation. The Pass 8 \textit{Fermi}-LAT $\gamma$-ray data of Mrk 180 was extracted from Fermi Science Support Center (FSSC) data server \citep{Fermi_Data_Download} for a period of more than 12.8 years (August 2008 to May 2021). We have used Fermipy (v1.0.1;\cite{2017ICRC...35..824W}), an open-source python package to analyze \textit{Fermi}-LAT $\gamma$-ray data. Moreover, we have used \textit{Fermi}-LAT Fourth Source Catalog Data Release 2 (4FGL-DR2; gll\_psc\_v27.fits) \citep{https://doi.org/10.48550/arxiv.2005.11208}. We have modeled the Galactic diffuse emission by the latest model template (gll\_iem\_v07; \cite{Acero_2016}) and for the extra-galactic isotropic diffuse emission model, we have considered iso\_P8R3\_SOURCE\_V2\_v1.txt. We have followed Fermipy's documentation for further analysis \citep{fermipy_doc} and extracted the light curve and SED of Mrk 180.
 
The photon-like events are classified as evclass=128. As the \textit{Fermi}-LAT collaboration recommended to use the ‘SOURCE’ event class for relatively small regions of interest ($<25^\circ$) \citep{bruel2018fermilat} and we have used the ‘P8R3 SOURCE’ event class for which ‘evclass’ has to be set to a value 128 \citep{Fermi-LAT_Ciceron} and evtype=3; each event class includes different event types which allows us to select events based on different criteria. The standard value of ‘evtype’ is 3 which includes all types of events i.e. front and back sections of the tracker (denoted by FRONT+BACK), for a given class. We have extracted the \textit{Fermi}-LAT $\gamma$-Ray data from FSSC data server considering a search radius of 30$^\circ$ around the source Mrk 180. During the data preparation, we have selected a ‘Region of Interest (ROI)’ of 10$^\circ,$ as suggested in Fermi’s Data Preparation page \footnote{\url{https://fermi.gsfc.nasa.gov/ssc/data/analysis/documentation/Cicerone/Cicerone_Data_Exploration/Data_preparation.html}} and the maximum zenith angle of 90$^\circ$ was chosen to avoid earth limb contamination in our analysis. We restricted our analysis to an energy range of 100 MeV to 500 GeV. We have obtained the $\gamma$-ray light curve shown in Fig.~\ref{fig:Mrk180_LC} and the SED, which is used to construct the multi-wavelength SED shown in Fig.~\ref{fig:Pure_LP_MWSED}, ~\ref{fig:Lp_UHECR_MWSED} and ~\ref{fig:LP_PP_MWSED}.


\subsection{\label{subsec:swift}SWIFT XRT and UVOT Data Analysis}
Neil Gehrels Swift observatory is a multi-wavelength space-based observatory with three instruments onboard: Burst Alert Telescope (BAT; 15.0- 150.0 keV), X-Ray Telescope (XRT; 0.3- 10.0 keV) and Ultraviolet and Optical Telescope (UVOT; 170- 600 nm) \citep{Burrows_2005}. It observes the sky in hard X-ray, soft X-ray, ultraviolet, and optical wavebands. Swift provides simultaneous data of any transient activity in all wavebands ranging from X-ray to optical. We collected all the XRT and UVOT data over the period August 2008 to May 2021, available for Mrk 180. We have analyzed 44 observations. The standard data reduction procedure\footnote{\url{https://www.swift.ac.uk/analysis/index.php}} has been followed to extract the source and background region.
\par
In Swift-XRT data, we have used clean event files corresponding to Photon-Count mode (PC mode), which we have obtained using a task ‘xrtpipeline’ version 0.13.5. The calibration file (CALDB), version 20190910, and other standard screening criteria have been applied to the cleaned data. A radius
of interest of 20-30 pixels has been considered to mark the source region, the radius of the background region is also the same, but it is far away from the source region. With the
help of ‘xselect’ tool, we have selected the source region and background region and saved the spectrum files of the corresponding regions. Then ‘xrtmkarf’ and ‘grppha’ tools have been used to generate ancillary response files (arfs) and group the spectrum files with the corresponding response matrix file (rmf); thereafter ‘addspec’ and ‘mathpha’ have been used. Thus we have obtained the spectrum. Thereafter, the spectrum has been modeled with xspec (v12.11.0; \cite{1996ASPC..101...17A} ) tools. We have included the absorption by neutral hydrogen having column density, N$_{\rm H}$= 1.37$\times10^{20}$ cm$^{-2}$ \footnote{\url{https://heasarc.gsfc.nasa.gov/cgi-bin/Tools/w3nh/w3nh.pl}} \citep{Bekhti_2016}. The final X-ray SED obtained in this way has been shown in Fig.~\ref{fig:Pure_LP_MWSED}, ~\ref{fig:Lp_UHECR_MWSED} and ~\ref{fig:LP_PP_MWSED}.

Mrk 180 was also monitored by Swift UVOT in all six filters: U(3465 \AA), V (5468 \AA), B (4392 \AA), UVW1 (2600 \AA), UVM2 (2246 \AA) and UVW2 (1928 \AA). The source region has been extracted from a region of 5" around the source, keeping the source at the center of the circle. The background region has been taken $\sim$3 times larger than the source region and it is far away from the source region. Using ‘uvotsource’ tool, we have extracted
the source magnitude. This magnitude does not include the galactic absorption, so it has been corrected. A python module ‘extinction’ \citep{Extinction} has been used to get the extinction values corresponding to all the Swift-UVOT filters. We have considered Fitzpatrick (1999) (Fitzpatrick 1999) dust extinction function for R$_{\rm V}$=3.1, where R$_{\rm V}$ is a dimensionless quantity, which is the slope of the extinction curve. For diffused interstellar medium (ISM) the mean value of R$_{\rm V}$ is 3.1 \citep{1975A&A....43..133S, 1985ApJ...288..618R, 1980MNRAS.192..467W}. Following are the values of the extinction coefficients of different Swift-UVOT wavebands which we have used in this work; U: 0.05584, V: 0.03460, B: 0.04603, UVW1: 0.07462, UVM2: 0.10383, UVW2: 0.09176.


\subsection{\label{subsec:xmm}XMM-Newton X-ray Data Analysis}

XMM-Newton is a space-borne X-ray observatory, consisting of three imaging X-ray cameras (European Photon Imaging Camera or EPIC), two grating X-ray spectrometers (Reflection Grating Spectrometer or RGS) and one optical monitor (OM). It was launched on December 10, 1999. Because of its great capacity to detect X-rays, it was formally known as the High Throughput X-ray Spectroscopy Mission. Now it is called XMM because of its multi-mirror design. The three EPIC cameras are the primary instrument aboard XMM-Newton; out of the three, two of them are MOS-CCD cameras and the remaining one is pn-CCD camera. The energy range of EPIC is about 0.15 keV- 15.0 keV. The MOS-CCD cameras are used to detect low-energy X-rays, whereas pn-CCD camera is used to detect high-energy X-rays. RGS operates from 0.35 keV to 2.1 keV. OM covers from 170 nm to 650 nm. From the data archive of XMM-Newton \footnote{\url{http://nxsa.esac.esa.int/nxsa-web/##search}}, we found two observations for Mrk 180: 0094170101 and 0094170301 of 20 ks and 8 ks respectively. We have followed standard data reduction procedure \footnote{\url{https://www.cosmos.esa.int/web/xmm-newton/sas-threads}} to extract the SED. We have extracted SED points from MOS1 and MOS2; combined them and finally, we got SED points from MOS. Also, we extracted SED points from pn detector. Thereafter, we used xspec  (v12.11.0; \cite{1996ASPC..101...17A}) to model these spectra. Other than X-ray data, we have also analyzed OM image mode data. Following the same data reduction procedure, we prepared the data; then we used `omichain' for further analysis. We followed \cite{OMICHAIN} instruction for the last step. By the `om2pha' \citep{om2pha} command, we extracted the spectrum file to analyze in xspec. For this step, the required OM response files have been copied from \cite{OMResponseFile}. The first observation 0094170101, contains single data corresponding to the u-band, which is insufficient for further analysis whereas, the second observation 0094170301 does not contain any image file for further study. So, our multi-wavelength data does not contain any XMM-Newton OM data.


\subsection{\label{subsec:mojave}MOJAVE Data}
	
 MOJAVE is a long-term program to monitor radio brightness and polarization variation in jets associated with active galaxies visible in the northern sky \citep{MOJAVEWebpage}. MOJAVE observes at three wavelengths 7 mm, 1.3 cm, and 2 cm, to obtain a full polarization image with an angular resolution better than 1 millisecond. We have collected MOJAVE data for Mrk 180 from the MOJAVE/2cm Survey Data Archive \citep{MOJAVESourcePage}. There are seven observations in the archive and used those data to construct the multi-wavelength SED.


\subsection{\label{subsec:magic}MAGIC Data}
 MAGIC is a system of two Imaging Atmospheric Cherenkov Telescopes (IACT), situated on the Canary Island of La Palma. 
VHE $\gamma$-rays impinging the earth's upper atmosphere, initiate cascade interactions; leading to the production of a shower of secondary particles, mainly electrons and positrons. Electrons and positrons moving faster than the phase velocity of light in the atmosphere emit Cherenkov radiation mainly in the UV-blue band for a duration of a few nanoseconds. MAGIC collects the Cherenkov light and focuses it onto a pixelized camera, composed of 576 photomultipliers (PMTs). Using dedicated image reconstruction algorithms, the energy and incoming direction of the primary $\gamma$-ray are calculated \citep{Doro2008}. This telescope can detect $\gamma$-rays of energy between 30 GeV to 100 TeV. VHE $\gamma$-rays from Mrk 180 were detected during an optical outburst in 2006 \citep{2006ApJ...648L.105A}. We have used that data from \citet{MAGIC} for our study.


\subsection{\label{subsec:archival}Archival Data}
We have collected the archival data from SSDC SED builder \citep{SSDC} and shown it with grey squares in the multi-wavelength SEDs (Fig.~\ref{fig:Pure_LP_MWSED}, ~\ref{fig:Lp_UHECR_MWSED} and ~\ref{fig:LP_PP_MWSED}).


\section{\label{subsec:light_curve}Fermi-LAT gamma-Ray Light curve Analysis}

\begin{figure}[h]
    \includegraphics[width=0.48\textwidth]{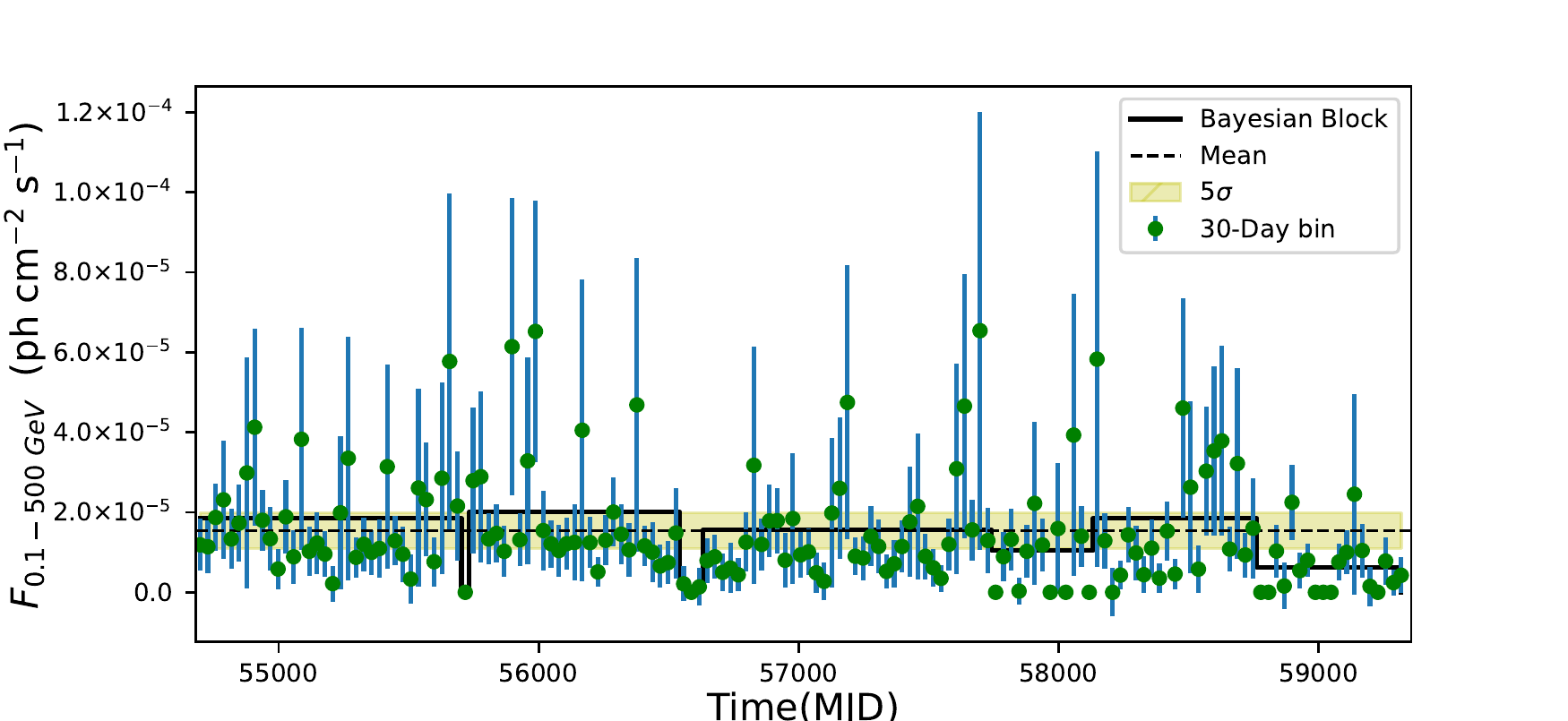}
    \caption{Application of Bayesian Block Method on \textit{Fermi}-LAT $\gamma$-Ray Data of Mrk 180 (MJD 54682.65- 59355.67)}
    \label{fig:Mrk180_LC}
\end{figure}

We analyzed 12.8 years (MJD 54682.65-59355.67) of \textit{Fermi}-LAT $\gamma$-ray data. Fig.~\ref{fig:Mrk180_LC} is the 30-day binned \textit{Fermi}-LAT $\gamma$-ray light curve. We have used the Bayesian Block method \citep{scargle2013bayesian} to detect any fluctuations. We have not found any significant variation in the $\gamma$-ray flux. Though there are a few data points with high $\gamma$-ray flux, those points have large error bars, so further analysis with a smaller bin size is not feasible in this case. We proceed to build up the SED with the long-term data, as this source does not have any obvious temporal features.


\section{\label{sec:sed_model}Multi-Wavelength SED Modeling}
\begin{figure*}
    \centering
    \includegraphics[width=0.75\textwidth]{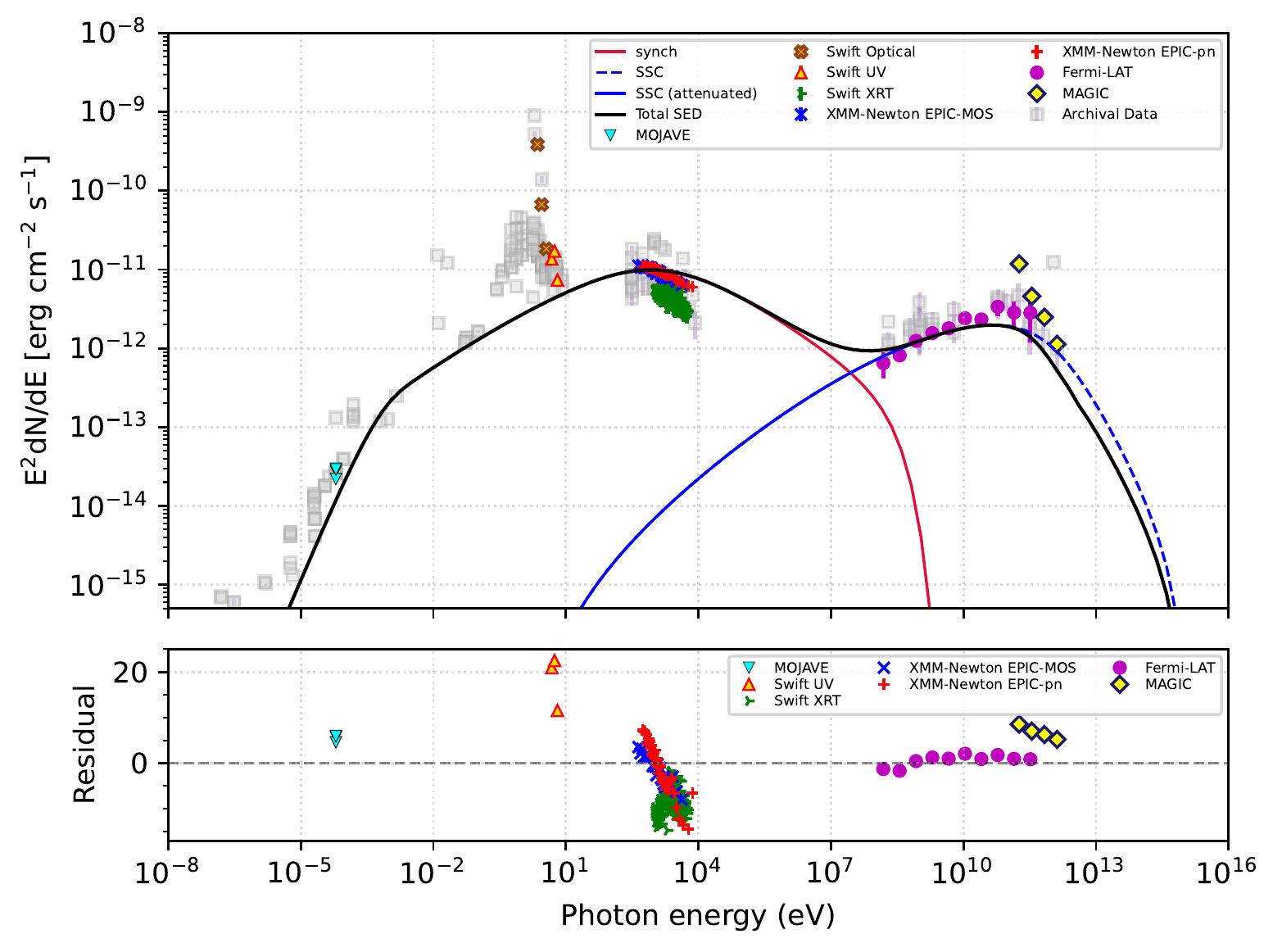}
    \caption{Pure leptonic modeling of multi-wavelength SED of Mrk 180 and residual plot corresponding to this modeling. The data color codes are mentioned in the plots.}
    \label{fig:Pure_LP_MWSED}
\end{figure*}

 \textit{Fermi}-LAT $\gamma$-ray, Swift X-ray, ultraviolet \& optical data and XMM-Newton X-ray data have been analyzed and archival data from MOJAVE, MAGIC and SSDC have been compiled to plot the SED covering radio to VHE $\gamma$-ray frequencies. As discussed previously, Fig.~\ref{fig:Pure_LP_MWSED}, ~\ref{fig:Lp_UHECR_MWSED} and ~\ref{fig:LP_PP_MWSED} shows the double hump structure of the blazar SEDs. We have modeled the SED using pure leptonic and lepto-hadronic scenarios. For the latter, we consider the 
 the line of sight component of the electromagnetic cascade, initiated by UHECR interactions \citep{Essey_10a, Essey_10b}, and also 
 $pp$ interaction as the origin of VHE $\gamma$-rays. 
 An external radiation field is required to
  produce a significant flux of secondary $\gamma$-rays in p$\gamma$ interactions, hence we do not include this scenario in this work.
   In the following subsections, we discuss about the models used in this work to explain the multi-wavelength SED of Mrk 180.


\subsection{\label{subsec:leptonic}Leptonic Modeling}
We have considered a spherical emission region of radius $R$ within the jet, moving with a Doppler factor $\delta_D$, where relativistic electrons and positrons, accelerated in the jet lose energy through synchrotron radiation in a steady and uniform magnetic field $B$, and also by SSC emission. From the maximum likelihood analysis of \textit{Fermi}-LAT data, a log-parabola injection was found to best fit the data. \cite{Massaro_2003} showed that a log-parabolic photon spectrum can be produced from the radiative loss of a log-parabolic electron spectrum. So, we have used the log-parabolic spectrum of the injected electrons in the blob to explain the multi-wavelength SED of Mrk 180, given by the following expression,
\begin{equation}
    Q(E)=L_0(E/E_0)^{-(\alpha+\beta \log_{10}(E/E_0))}
    \label{eqn:leptonic_injection}
\end{equation}
where $Q(E)$ is the log-parabolic distribution, $L_0$ is the normalization constant, $E_0$ is the scaling factor or pivot energy which is set to 97 MeV in our modeling and kept fixed, $\alpha$ is the spectral index and $\beta$ is the curvature index.

We have used the open-source code \cite{libgamera} \citep{Hahn:2015hhw} to model the multi-wavelength leptonic emission. It solves the time-dependent transport equation and propagates the particle spectrum $N(E,t)$ for an injected spectrum $Q(E)$ to calculate the synchrotron and SSC emissions including the Klein-Nishina effect. GAMERA solves the following transport equation,
\begin{equation}
    \dfrac{\partial N(E,t)}{\partial t}= Q(E)-\dfrac{\partial}{\partial E}(b(E,t)N(E,t))-\dfrac{N(E,t)}{\tau_{\rm esc}}
\end{equation}
where, $Q(E)$ is the input particle spectrum, $b(E,t)$ corresponds to the energy loss rate by synchrotron and SSC emission. The term $\tau_{\rm esc}(E,t)$ denotes the escape time of particles from the emission region. We consider a constant escape of the electrons from the emission region  over the dynamical timescale, $\tau_{\rm esc}\sim R/c$, where $c$ is the speed of light. We find that the time-evolved electron spectrum reaches the steady state after nearly 100 days, and this spectrum has been used in this work. 

\subsection{\label{subsec:los_uhecr}UHECR interactions}

We have assumed a power-law injection of the protons into the interstellar medium (ISM) of the following form
\begin{equation}
	    N_p(E_p)=\frac{dN}{dE_p}=A_p E_p^{-\alpha_p}
	    \label{eqn:UHECR_injection}
\end{equation}
where $A_p$ is the normalization constant of the injected proton spectrum, $\alpha_p$ is the spectral index, which is the same for electrons and protons as they are accelerated in the same region. We have taken the minimum energy of protons E$_{\rm p, min}$= 0.1 EeV, and the maximum energy of protons E$_{\rm p, max}$= 100 EeV. 

The ultra-high energy protons escape from the emission region and propagate through the extra-galactic medium interacting with CMB and EBL photons.  In this process, electrons, positrons, $\gamma$-rays, and neutrinos are produced through $\Delta$-resonance and Bethe-Heitler pair production. Protons interact with the CMB and EBL photons in the following way,
\begin{equation}
    p+ \gamma_{bg}= p+ e^+ + e^-
\end{equation}
\begin{equation}
    p+ \gamma_{bg}\rightarrow \Delta^+ \rightarrow 
    \begin{cases}
        n+ \pi^+ \\
        p+ \pi^\circ
	\end{cases}
\end{equation}
The neutral pions decay to gamma photons ($\pi^\circ \rightarrow \gamma \gamma$) and the charged pions decay to neutrino ($\pi^+ \rightarrow \mu^+ + \nu_\mu \rightarrow e^+ + \nu_e + \bar{\nu}_\mu + \nu_\mu $). The resulting cosmogenic neutrinos propagate undeflected by magnetic fields and unattenuated by interaction with other particles.

The secondary $e^\pm$, $\gamma$-rays initiate electromagnetic (EM) cascade by undergoing pair production, inverse-Compton upscattering of the background photons, and synchrotron radiation in the extragalactic magnetic field (EGMF). The resulting spectrum extends down to GeV energies and depends more on the propagation distance and background photon model than the injection parameters. We use the semi-analytical EBL model given in \cite{Gilmore:2011ks} for the propagation of UHECR and the attenuation of secondary EM particles, and also the primary $\gamma$-rays coming from leptonic emission inside the source. UHECRs also interact with the universal radio background \citep{Protheroe:1996si} which is important at energies higher than the Greisen-Zatsepin-Kuzmin (GZK) cutoff energy for $\Delta$-resonance with the CMB photons. The EGMF causes a spreading of the UHECR beam and also the EM particles. We consider the contribution from the line of sight resolved component of the cascade spectrum to the observed SED (cf. Sec.~\ref{sec:results}). 

We have used the publicly available simulation framework, \textsc{CRPropa 3} \citep{AlvesBatista:2016vpy, AlvesBatista:2022vem, Heiter_18} to propagate UHECR protons from their source to the observer. The secondary EM particles are propagated in the CRPropa simulation chain, using a value of EM thinning $\eta=0.6$.

\begin{figure*}
    \centering
        \includegraphics[width=0.75\textwidth]{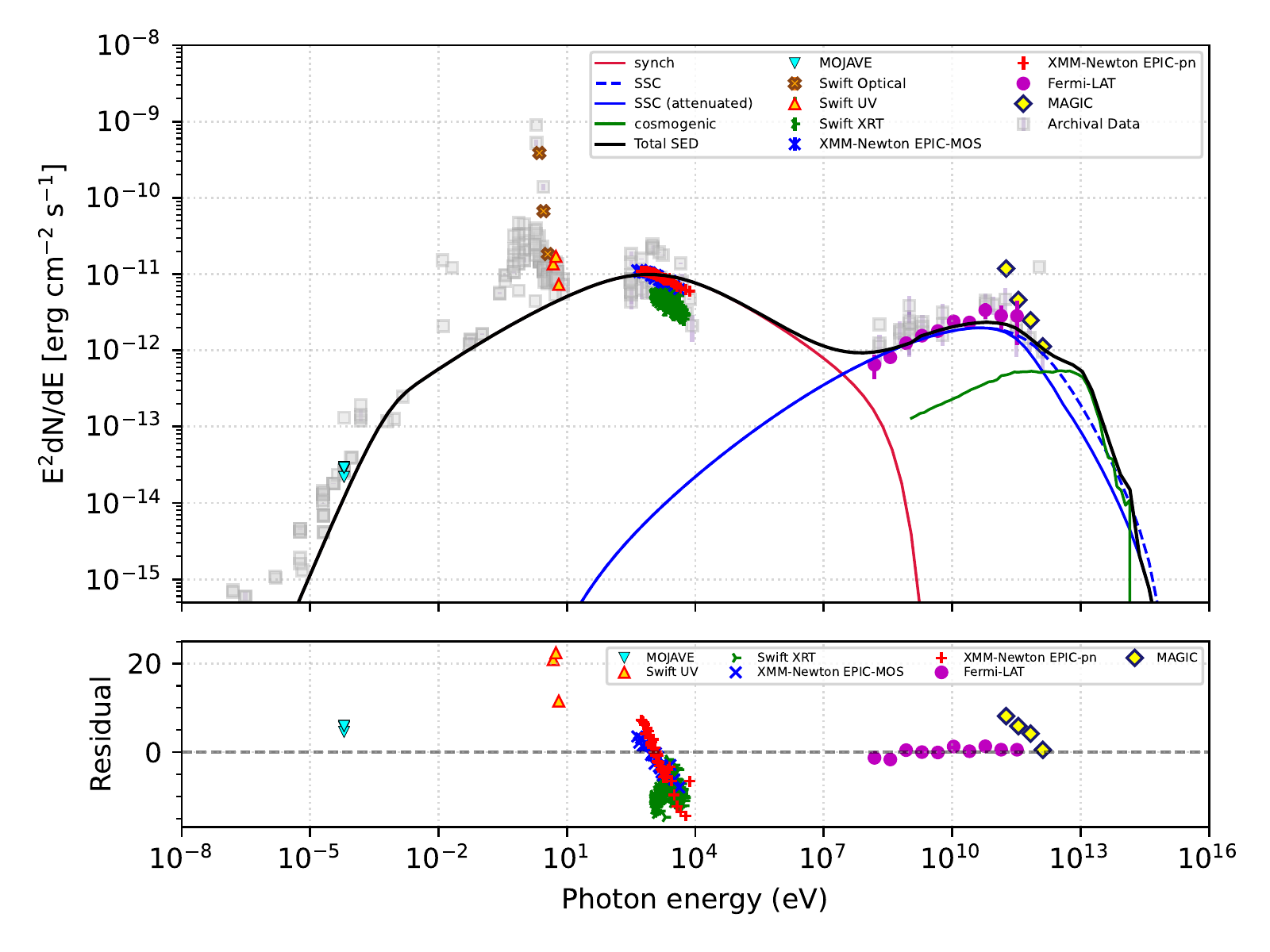}
     \caption{Leptonic+ hadronic (UHECR) modeling of multi-wavelength SED of Mrk 180 and residual plot corresponding to this modeling.}
     \label{fig:Lp_UHECR_MWSED}
\end{figure*}

\subsection{\label{subsec:pp} pp interactions}
An alternative scenario is when the relativistic protons have much lower energy than UHECRs and they interact with the cold protons within the emission region as they are trapped in the magnetic field of the emission region. The proton-proton interactions result in the production of neutral and charged pions. These pions decay into secondary particles e.g. electrons/ positrons, neutrinos and $\gamma$-rays. The proton-proton interaction channels can be shown in the following manner:
\begin{align}
p+p\rightarrow 
    \begin{cases}
    \pi^\circ \rightarrow  \gamma+\gamma\\
    \pi^+ \rightarrow \nu_\mu + \mu^+ \rightarrow \nu_\mu + e^+ + \nu_e + \bar{\nu_\mu} \\
    \pi^- \rightarrow \bar{\nu_\mu} + \mu^- \rightarrow \bar{\nu_\mu} + e^- + \bar{\nu_e} + \nu_\mu
    \end{cases}
\end{align}

We have considered a power-law proton injection spectrum within the emission region, with a spectral index $\alpha_p$  and minimum E$_{\rm p, min}$ and maximum energy E$_{\rm p, max}$. We have used the publicly available code \textsc{Gamera} for the time-independent $pp$ modeling. It uses the formalism given in \citet{2014PhRvD..90l3014K}. There are four hadronic interaction models that are included in this code, and for our work, we have used the one given by \textsc{Pythia 8.18} \citep{Sj_strand_2008}. 

We have balanced the total charge in the emission region to determine the total number of protons. The $\gamma$-ray spectrum produced in $pp$ interactions has been corrected for internal absorption by the lower energy photons inside the blob, and also for absorption by the EBL.

\begin{figure*}
    \centering
    \includegraphics[width=0.75\textwidth]{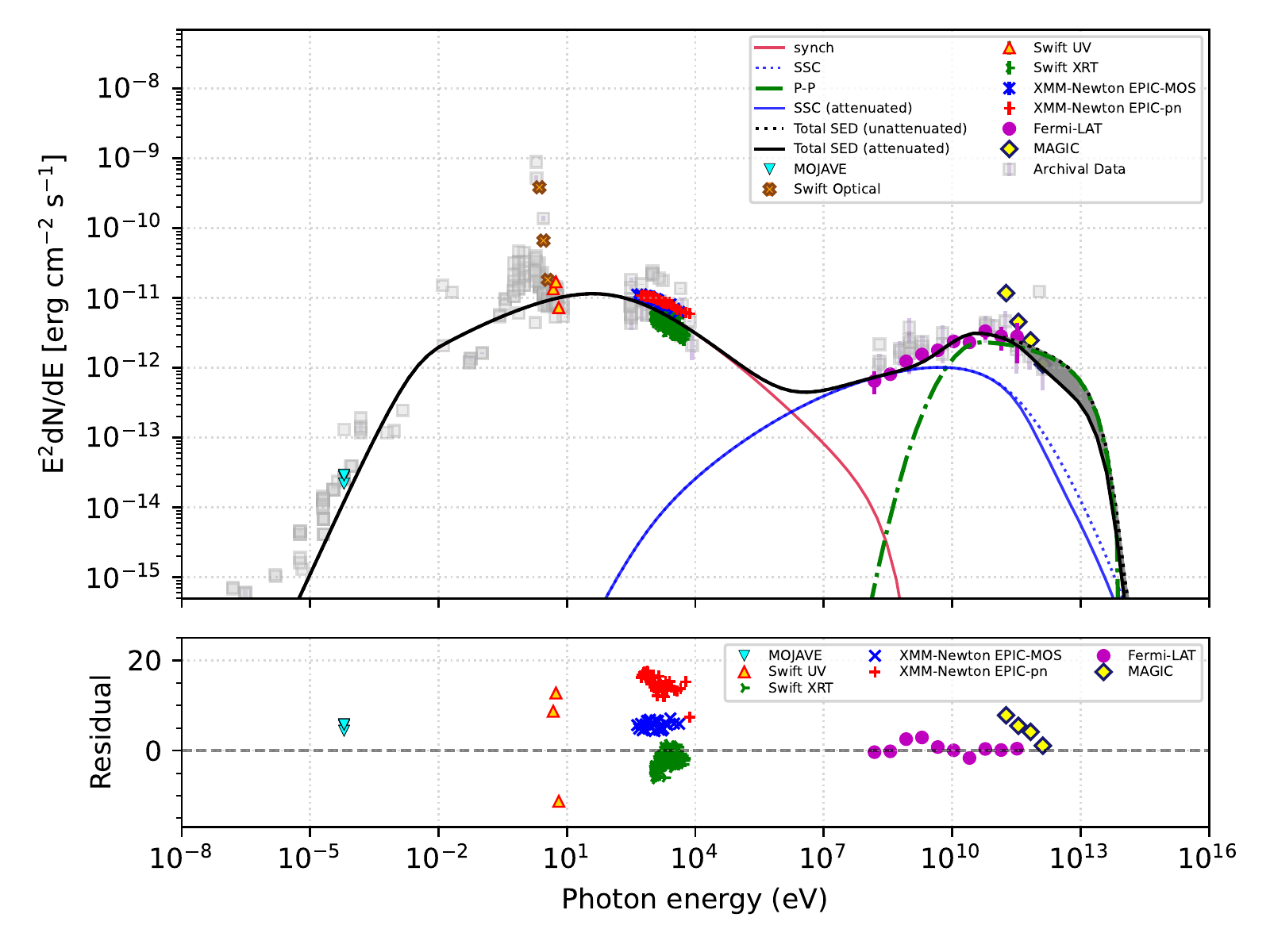}
        \caption{Leptonic+ hadronic ($pp$) modeling of multi-wavelength SED of Mrk 180 and residual plot corresponding to this modeling; the grey-shaded region denotes the difference between the attenuated and unattenuated regions of the total SED.}
    \label{fig:LP_PP_MWSED}
\end{figure*}

\subsection{\label{subsec:power}Jet Power}
We have calculated the kinematic jet power using the following equation
\begin{equation}
    P^k_{\rm tot}= P_e+ P_B + P_p =\pi R^2 \Gamma^2 c (u'_e+u'_p+u'_B)
    \label{eqn:JetPow}
\end{equation}

where P$^k_{\rm tot}$ is the kinematic jet power, $\Gamma$ is the bulk Lorentz factor; $u'_e$, $u'_p$ and $u'_B$ are the energy densities of the relativistic electrons (and positrons) and protons and magnetic field respectively in the comoving jet frame \citep{PhysRevD.99.103006, PhysRevD.101.063024}.The primed and unprimed notations denote quantities in the comoving jet frame and the AGN frame, respectively.
We have maintained the charge neutrality condition in the jet. 
If we add the jet power of cold protons the luminosity budget in proton-proton interaction model exceeds the Eddington luminosity as discussed in \citep{PhysRevD.99.103006, PhysRevD.101.063024}. A sub-Eddington jet power in proton-proton interaction model is possible in the scenario discussed in a recent paper \citep{2022PhRvD.106j3021X} after including the jet power in cold protons. However, we compare only the kinematic jet power to the Eddington luminosity as it has been done in earlier papers. 

Here, we have considered, the bulk Lorentz factor ($\Gamma$) and Doppler factor ($\delta_D$) are equal. We have presented the jet powers  of individual components and the total kinematic jet power in Table \ref{tab:GAMERA_Fitting_param}.

The mass of the black hole of Mrk 180 as reported in earlier papers has been used to calculate the Eddington luminosity. According to \cite{2003ASPC..290..621T}, the value of $\log_{10}(M_{\rm BH}/M_\odot)$ is 8.59 and according to \cite{2003mglh.conf..109F}, the value of $\log_{10}(M_{\rm BH}/M_\odot)$ is 8.70; where $M_{\rm BH}$ is the mass of the black hole and $M_\odot$ is the solar mass. Using these values, we have calculated the Eddington luminosity (L$_{\rm Edd}$) of Mrk 180, which are 5.06$\times10^{46}$ erg/s and 6.51$\times10^{46}$ erg/s respectively. The total kinematic jet powers obtained in our models are less than the Eddington luminosity of Mrk 180.

\section{\label{sec:results}Results}

Mrk 180/ Mkn 180/ TeV J1136+701 or 4FGL J1136.4+7009 is an HBL type blazar at a redshift of 0.045. This source is monitored by several telescopes viz. \textit{Fermi}-LAT, Swift, XMM-Newton, MOJAVE, MAGIC, KVA, ASM, RATAN-600, Mets\"{a}hovi, Effelsberg, IRAM throughout the year, and it was closely monitored during the high state in optical waveband in 2006. 

12.8 years (MJD 54682.65- 59335.67) of \textit{Fermi}-LAT $\gamma$-ray data of Mrk 180 has been analyzed in this work. Besides \textit{Fermi}-LAT $\gamma$-ray data, we also collected data in other wavebands e.g. Swift, XMM-Newton, MOJAVE, and MAGIC. Fig.~\ref{fig:Mrk180_LC} is the long-term \textit{Fermi}-LAT $\gamma$-ray light curve in 30-day binning. As can be seen from Sec.~\ref{subsec:light_curve}, this long-term light curve does not show any significant flaring throughout this time, also the error bars of the high-energy $\gamma$-ray data points are large, hence a more detailed analysis of the light curve cannot give us any useful information. To know about the physical processes which can explain the observed spectrum, we studied the long-term SED of Mrk 180; where we have used multi-wavelength data from different telescopes. The multi-wavelength SED shows the double hump structure, which has been modeled with GAMERA; considering a simple one-zone spherical emission region within the jet. In Fig.~\ref{fig:Pure_LP_MWSED}, ~\ref{fig:Lp_UHECR_MWSED} and ~\ref{fig:LP_PP_MWSED}, we have shown the multi-wavelength SEDs fitted with different models e.g. pure leptonic, lepto-hadronic. Also, we have shown the residual (Data-Model/error) plot corresponding to the fit to each model in Fig.~\ref{fig:Pure_LP_MWSED}, ~\ref{fig:Lp_UHECR_MWSED} and ~\ref{fig:LP_PP_MWSED}.

First, we consider a pure leptonic model (Fig.~\ref{fig:Pure_LP_MWSED}), where the first hump is produced due to the synchrotron radiation of the relativistic electrons, and the second hump is produced due to the up-scattering of the synchrotron photons by the relativistic electrons. As discussed in Sec.~\ref{subsec:leptonic}, we consider a spherical emission region or blob of radius R within the blazar jet. Leptons are injected within the blob following an injection spectrum (Eqn. \ref{eqn:leptonic_injection}). The best-fitted parameter values corresponding to this modeling e.g. spectral index ($\alpha$), curvature index ($\beta$) are listed in the first column of Table~\ref{tab:GAMERA_Fitting_param}. We have mentioned the jet power of different components e.g. relativistic leptons (P$_e$), magnetic field (P$_B$), and relativistic protons (P$_p$) in the Table~\ref{tab:GAMERA_Fitting_param}, also the total kinematic jet power (P$^k_{\rm tot}$) which is the sum of the jet power of all the components of a model. 

\begin{figure}
    \includegraphics[width=0.48\textwidth]{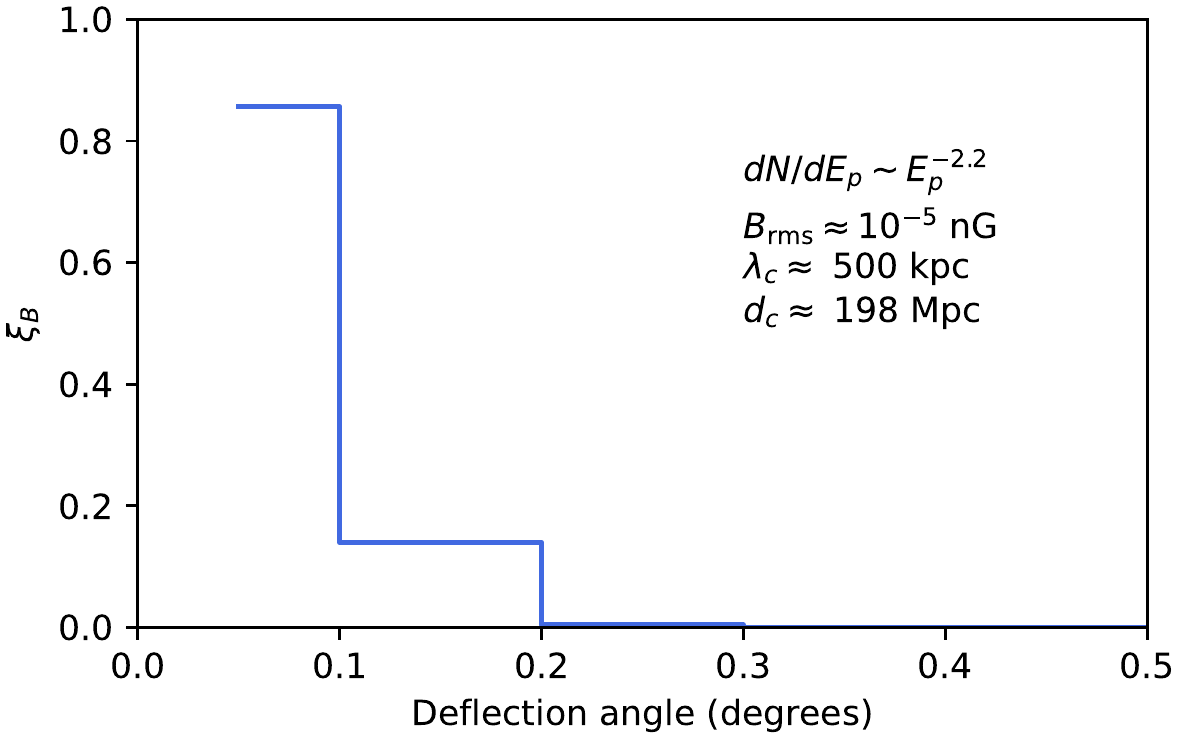}
    \caption{Distribution of propagated UHECRs as a function of deflection angle in a random turbulent magnetic field.}
    \label{fig:UHECR_Distribution}
\end{figure}

\begin{figure*}
    \centering
    \includegraphics[width=0.32\textwidth]{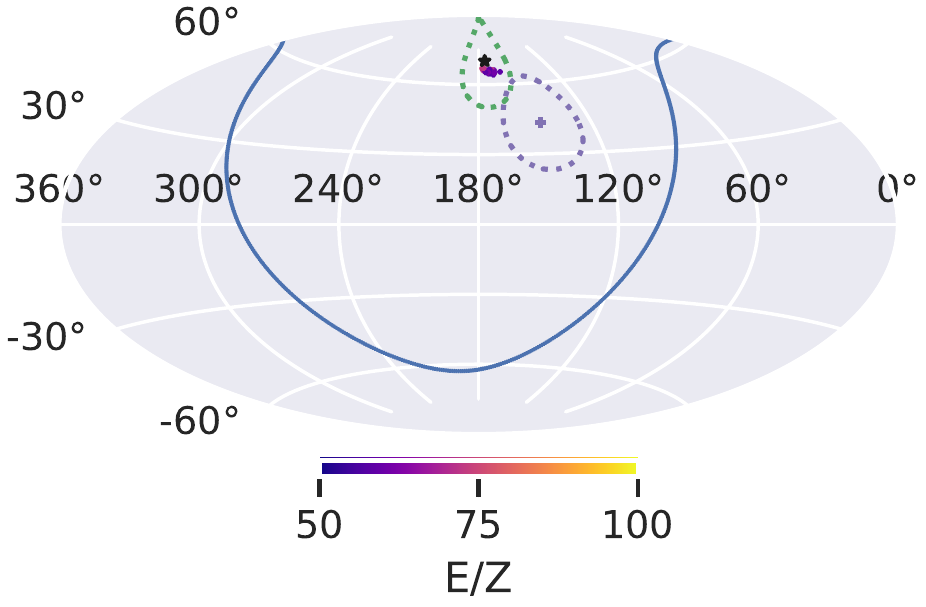}%
    \includegraphics[width=0.32\textwidth]{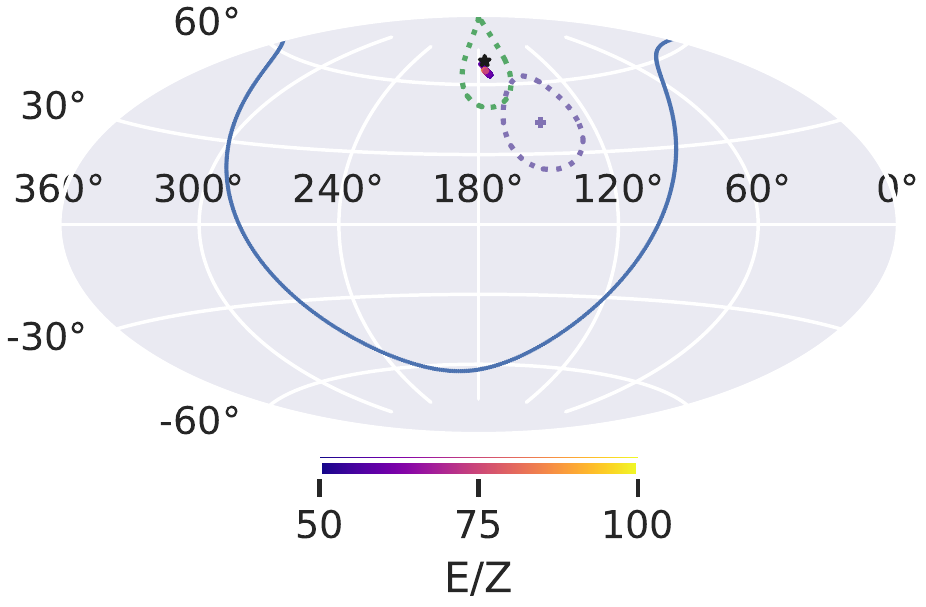}%
    \includegraphics[width=0.32\textwidth]{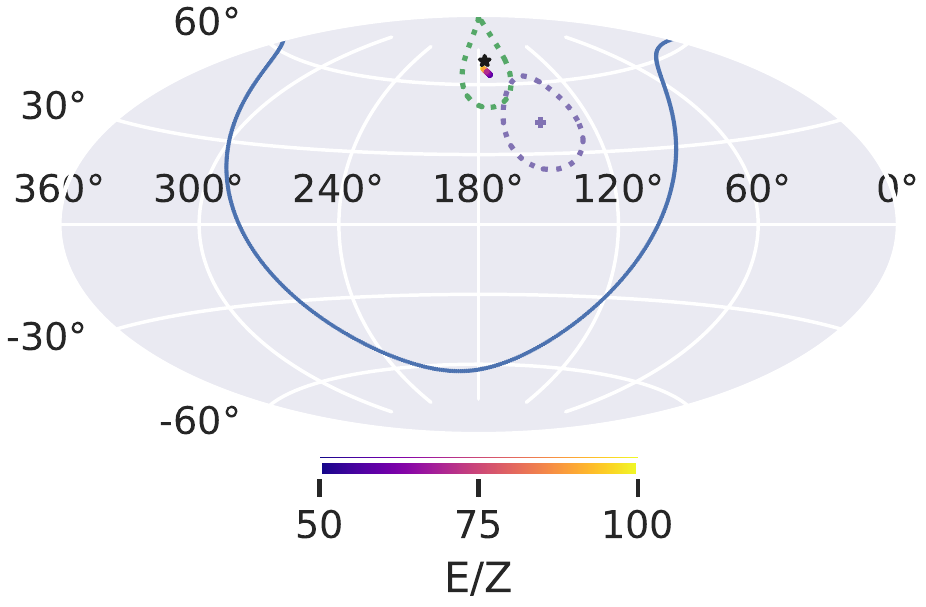}
    \caption{Arrival direction of UHECRs at $E>57$ EeV from Mrk 180 to Earth. The blue line shows the Galactic plane. The purple point and the purple dotted curve show the TA hotspot center and the 20$^\circ$ region around it. Similarly, the green dotted curve shows the $20^\circ$ region around Mrk 180. The color bar indicates the energy per nucleon (E/z) of the observed events. From left, the figures correspond to (a) pure proton injection and B$_{\rm rms }\approx10^{-3}$ nG; (b) pure proton injection and B$_{\rm rms}\approx10^{-5}$ nG; (c) Fe injection and B$_{\rm rms}\approx10^{-5}$ nG}. 
    \label{fig:aniso}
\end{figure*}

The pure leptonic model is found to be insufficient to explain the multi-wavelength SED, as the highest energy $\gamma$-ray data point cannot be fitted with this model. Moreover, the slope of the observed X-ray spectrum does not match the slope of the synchrotron spectrum obtained in our model.  To improve the fit to the multi-wavelength SED, particularly at the VHE $\gamma$-ray regime, we check the fit with lepto-hadronic models. As discussed earlier, we have considered two kinds of hadronic processes, viz., the UHECR interaction with the background photons and the $pp$ interaction within the blob.

In the case of UHECRs (for simplicity we consider only protons), the escape of protons from the blazar jet can dominate over the energy loss inside the blazar jet. We consider a power-law injection of protons into the interstellar medium (ISM) following Eqn.~\ref{eqn:UHECR_injection}. We have considered proton injection into ISM between E$_{p, \rm min}$= 0.1 EeV and E$_{p, \rm max}=100$ EeV. The injection spectral index $\alpha_p=2.2$ is the same as for leptons. In the UHECR interaction model, we consider the three-dimensional propagation of UHECRs to calculate the fraction of them that survives within $0.1^\circ$ degrees of initial emission direction and denote it by $\xi_B$. Protons are propagated from the source at a comoving distance $\sim200$ Mpc and collected over a spherical region of radius 1 Mpc. We consider a random turbulent EGMF given by a Kolmogorov power spectrum and an RMS field strength of B$_{\rm rms}\approx10^{-5}$ nG and a coherence length of 0.5 Mpc using wave modes between 80 kpc and 2.25 Mpc. The distribution of survival fraction with deflection angle is shown in Fig.~\ref{fig:UHECR_Distribution}. We multiply the flux of cosmogenic $\gamma$-ray spectrum by $\xi_B$ to take into account the $\gamma$-rays reaching the observer from the direction of the blazar. The \textit{Fermi}-LAT resolution to a single photon above 10 GeV is $\sim 0.15^\circ$.

Fig.~\ref{fig:Lp_UHECR_MWSED} is the resulting fit corresponding to this model. The green curve indicates the spectrum of cosmogenic photons. The required power in UHECR protons is calculated in the following manner \citep{Das_2020},
\begin{equation}
    P_{\rm UHECR} = \dfrac{2\pi d_L^2(1-\cos\theta_{\rm jet})}{\xi_B f_{\rm CR}} \int_{\epsilon_{\gamma,\rm min}}^{\epsilon_{\gamma,\rm max}} \epsilon_\gamma\dfrac{dN}{d\epsilon_\gamma dAdt}d\epsilon_\gamma \label{eqn:UHECR_Luminosity}
\end{equation}
where d$_L$ is the luminosity distance of Mrk 180, $\theta_{\rm jet}$ is jet opening angle, $\xi_B$ is the the survival rate of UHECR within 0.1$^\circ$ of the direction of propagation to the observer. The quantity $f_{\rm CR}$ is the fraction of UHECR luminosity that goes into cosmogenic $\gamma$-rays and depends on the propagation distance. The integration is done over the cosmogenic photon spectrum allowed by the observed SED. d$_L$ is 207 Mpc,  $\theta_{\rm jet}$ is 0.1 radians (we have considered typical value of $\theta_{\rm jet}$ or jet opening angle) \citep{Finke_2019, refId0} and $\Gamma$=20. For the chosen parameters, $\xi_B=$ 0.85 and $f_{\rm CR}=0.03$. Putting these values in Eqn. \ref{eqn:UHECR_Luminosity}, P$_{\rm UHECR}$ has been calculated. Finally, we add up the total kinematic jet power of the relativistic leptons, magnetic field and UHECRs denoted by P$_{\rm UHECR}$ to get the total kinematic jet power for this model to be 2.9$\times10^{43}$ erg/s, which is less than the Eddington luminosity of Mrk 180 by several orders of magnitude. The best-fitted values of this model are tabulated in the second column of Table~\ref{tab:GAMERA_Fitting_param}.
In this case, the highest energy MAGIC data point can be fitted, but the fit to the X-ray data points has not improved.
\par 
We subsequently consider the $pp$ interactions within the jet. As explained in sec. \ref{subsec:pp}, the relativistically accelerated protons interact with cold protons and produced neutral and charged pions which decay into photons, leptons, and neutrinos. A power-law proton spectrum is injected within the blob with a spectral index $\alpha_p$=2.2, minimum (E$_{\rm p,min}$) and maximum energy (E$_{\rm p,max}$) 10 GeV and  10$^4$ GeV respectively and the cold proton density is assumed to be n$_{\rm H}$=1.2$\times10^6$ cm$^{-3}$.  These parameter values have been presented in the third column of Table~\ref{tab:GAMERA_Fitting_param}. Previously, \citet{PhysRevD.99.103006} showed $pp$ interaction model can explain the observed high-energy $\gamma$-rays from the blazar TXS 0506+056 for n$_{\rm H}=1.68\times10^6$ cm$^{-3}$. \cite{AHARONIAN2000377} showed that high-energy $\gamma$-ray production in an AGN jet via $pp$ interaction demands high cold proton density. To interpret the reported TeV flares of Markarian 501 by $pp$ interactions, they showed n$_{\rm H}$ should exceed $10^6$ cm$^{-3}$.

 From Fig.~\ref{fig:Pure_LP_MWSED} we can see that the SED from the pure leptonic model cannot fit the Swift UV data points. The slope of the observed X-ray and the $\gamma$-ray data points cannot be explained with the slope of the theoretical SED, also it poorly fits the $\gamma$-ray data points. The residual plot corresponding to the pure leptonic model shows this model poorly fits the Swift UV data, X-ray, and MAGIC data.
 
 From Fig.~\ref{fig:Lp_UHECR_MWSED} it can be seen that UHECR interactions make the fit better for the MAGIC data points but the slope of the SED from this model does not match the slope of the X-ray data. Moreover, the Swift UV data cannot be fitted well with this model. The residual plot corresponding to this model looks almost the same as that of the pure leptonic model between 10$^{-5}$- 10$^{11}$ eV, except for the MAGIC data points.
 
 Fig.~\ref{fig:LP_PP_MWSED} shows improvement in both SED and the residuals. The SED fits the Swift UV data points, and matches the slope of the X-ray data and the $\gamma$-ray data. The residual plot corresponding to this model shows that the residuals for the Swift UV data points lie in between $\sim\pm10$, whereas they lie in between +10 to +20 in Fig.~\ref{fig:Pure_LP_MWSED} and Fig.~\ref{fig:Lp_UHECR_MWSED}. In Fig.~\ref{fig:LP_PP_MWSED}, the residuals for the Swift XRT \& XMM-Newton EPIC-MOS data lie within $\pm10$ and the XMM-Newton EPIC-pn data lie out of +10. In the previous two plots i.e. Fig.~\ref{fig:Pure_LP_MWSED} and Fig.~\ref{fig:Lp_UHECR_MWSED}, all the residuals for the X-ray data points lie within +10 to -20. It is clear from the residual plot that the SED is not very well fitted which is why we are getting large values of the residuals. We have not shown the residuals for the Swift Optical data points, as they cannot be fitted with any of these models. 
 Most of the $\gamma$-ray data points can be fitted in this model.
 The total kinematic jet power corresponding to each model is less than the Eddington luminosity of Mrk 180, which has been mentioned in Table \ref{tab:GAMERA_Fitting_param}.\\

\subsection{UHECRs from Mrk 180}

It has been proposed earlier \citep{PhysRevD.93.043011} that Mrk 180 may be a source contributing to the UHECR hotspot observed by the Telescope Array (TA) collaboration above 57 EeV. We propagate UHECRs from the source to the Earth in a random turbulent magnetic field given by the Kolmogorov power spectrum. We consider three different combinations of the RMS value of the EGMF (B$_{\rm rms}$) and composition at the source as shown in Fig.~\ref{fig:aniso}. The turbulence correlation length of the EGMF is taken to be ~0.5 Mpc. The Galactic magnetic field model (GMF) is considered to be the one given in \cite{Jansson:2012pc}. We inject cosmic rays with a generic power-law spectrum given by $dN/dE\sim E^{-2}$ and perform three-dimensional simulations including both GMF and EGMF in \textsc{CRPropa 3} \citep{AlvesBatista:2016vpy, AlvesBatista:2022vem}. We consider two cases of composition with extreme masses, viz., $^1$H and $^{56}$Fe. For pure proton injection, the magnetic rigidity is higher and the resulting deflection is low. 

We show the case of B$_{\rm rms}\sim 10^{-3}$ nG and $10^{-5}$ nG in the left and middle panel of Fig.~\ref{fig:aniso} for proton injection. For the same injected luminosity, the number of detected events in the former case is 35, while that for the latter increases by almost three orders of magnitude. It can be seen that even with Fe injection (cf. right panel in Fig.~\ref{fig:aniso}), the angular width of the source observed through UHECRs doesn't show a significant change; although, the observed energy spectrum is different. Due to the photo-disintegration of the nuclei traversing a comoving distance of $\sim200$ Mpc, the observed events at Earth for the energy range considered are all protons. Thus, it can be seen that for optimistic magnetic field values considered, the contribution of this source to the TA hotspot is disfavored, unless, very high magnetic fields $\mathcal{O}\sim1$ nG or higher are considered. Although a higher spread in the arrival direction is expected if the detection threshold is lowered, the Galactic magnetic field shadows the directional signatures. Thus, Mrk 180 may not be a plausible UHECR source for explaining the TA hotspot.

\section{\label{sec:discussion}Discussions}

Being at a redshift of 0.045, Mrk 180 is an interesting source to study the radiative mechanisms producing TeV $\gamma$-rays. VHE $\gamma$-ray emission from this source was detected by MAGIC in 2006 \citep{Albert_2006} followed by an enhanced optical state. This source has been monitored by several telescopes viz. \textit{Fermi}-LAT, Swift, XMM-Newton, MOJAVE, MAGIC, KVA, ASM, RATAN-600, Mets\"{a}hovi, Effelsberg, IRAM throughout the year. Previously, \cite{https://doi.org/10.48550/arxiv.1110.6341} and  \cite{https://doi.org/10.48550/arxiv.1109.6808} studied this source using multi-wavelength data.
They have discussed about the results of the multi-wavelength campaign in 2008 covering radio to TeV $\gamma$-ray observations.
At that time Mrk 180 was known to be a TeV $\gamma$-ray source detected by MAGIC only a couple of years back.
Their study reported the first multi-wavelength campaign on Mrk 180. Optical observation was carried out by KVA telescope simultaneously with TeV $\gamma$-ray observation with MAGIC. The radio observation was carried out with RATAN-600, Mets\"{a}hovi, Effelsburg and IRAM. 
Swift XRT detected flux variability in X-rays. In the same observation window of Swift XRT, Mets\"{a}hovi and AGILE could not detect this source. 
\textit{Fermi}-LAT light curve showed enhancement in $\gamma$-ray flux during the second flare.
\par
They tried to explain the simultaneous multi-wavelength SED of Mrk 180 by two models: (1) a one-zone SSC model (2) a self-consistent two-zone SSC model; they considered the injected electron spectrum as a broken power-law distribution. It can be seen in \cite{https://doi.org/10.48550/arxiv.1110.6341} that during the high state both the models cannot explain the multi-wavelength data properly.
The steep X-ray spectrum and high optical flux could not be explained simultaneously assuming they were produced in the same zone.
Moreover, in the two-zone SSC model the required value of the Doppler factor $\delta$ is very high.  
During the low X-ray state both the models can explain the SED for moderate values of parameters. 
\cite{Nilsson_2018} studied R-band long-term optical data (over a span of $\sim$10 years) of 31 northern blazars and Mrk 180 is one of them. They 
could not find any significant periodicity for this source.
The earlier multi-wavelength studies on Mrk 180 have been complemented in this work with more data analysis and theoretical modeling of the SED over a long period of observations.

  For the temporal study, we analyzed 12.8 years (MJD 54682.65- 59355.67) of \textit{Fermi}-LAT $\gamma$-ray data. Fig.~\ref{fig:Mrk180_LC} is the long-term \textit{Fermi}-LAT $\gamma$-ray light curve in 30-day bin. No $\gamma$-ray flux enhancement has been found from this long-term light curve, also the error bars of the high-energy $\gamma$-ray data points are large to carry on a detailed temporal study on this source. To know about the physical processes we studied the long-term SED of Mrk 180. For this study, we have used multi-wavelength data from MOJAVE, MAGIC, Swift, XMM-Newton, and \textit{Fermi}-LAT. The SED shows typically the double hump structure. We have modeled this multi-wavelength SED with GAMERA. We have considered a simple one-zone spherical emission region within the jet. In Fig.~\ref{fig:Pure_LP_MWSED}, ~\ref{fig:Lp_UHECR_MWSED} and ~\ref{fig:LP_PP_MWSED}, we have shown the modeled multi-wavelength SEDs with different models e.g. pure leptonic, lepto-hadronic. Also, we have shown the residual plots of each model, attached just below that particular SED. The results of the multi-wavelength SED modeling with different models have already been discussed in Sec.~\ref{sec:results}. The leptonic modeling is not sufficient to explain the multi-wavelength SED of Mrk 180. 
  We have considered two lepto-hadronic models to improve the fit to the observed data points. The first model involves interactions of UHECRs injected by Mrk 180 with the radiation backgrounds, and in the second model, we have considered interactions of relativistic protons in the jet with the cold protons. The latter gives a slightly better fit to the data, however, more observational data is necessary to explain the radiation mechanisms in Mrk 180, as our results show large values of residuals in all the cases. We look forward to future multi-wavelength campaigns to cover all the frequencies over a long time period to monitor this source more closely.
  
\par
 \cite{PhysRevD.93.043011} calculated the probability associated with some sources to be the contributors to the TA hotspot, Mrk 180 is one of them. It is important to know the role of Mrk 180 as a UHECR accelerator, and whether it can generate events above 57 EeV. In our study for conservative values of EGMF, Mrk 180 is disfavoured as a source of the UHECR events contributing to the TA hotspot. In future, with more observational data it would be interesting to study the association of Mrk 180 with the TA hotspot.

 \section{\label{sec:conclusion}Conclusion}
The HBL Mrk 180, at a redshift of 0.045, is an interesting source to study the emission covering radio to VHE $\gamma$-ray frequency. We have analyzed the \textit{Fermi}-LAT $\gamma$-ray data detected from this source over a period of 12.8 years. The light curve analysis does not show any significant variation in flux. We have studied the long-term multi-wavelength SED of this source to understand the physical processes which can explain the HBL nature of this source. We modeled the multi-wavelength SED with a time-dependent code `GAMERA'. It is found that a single-zone pure leptonic model cannot explain the multi-wavelength spectrum of Mrk 180 properly. We considered single-zone lepto-hadronic models to obtain better fits to the data. 
The residuals of the three models are compared and the $pp$ interaction model is found to give a better fit to the multi-wavelength data compared to the other two models. 
More observational data covering the radio to VHE $\gamma$-ray
 frequency would be useful to explore the emission mechanisms of Mrk 180 and to give a definitive conclusion.
 The possible association of Mrk 180 with the TA hotspot events above 57 EeV has also been examined using the simulation framework \textsc{CRPropa 3} \citep{AlvesBatista:2016vpy, AlvesBatista:2022vem}. In this study we do not find any UHECR event from Mrk 180 contributing to the TA hotspot, hence, we conclude that for conservative values of EGMF, Mrk 180 is disfavoured as a source contributing to the TA hotspot, however, in future with more UHECR data it would be possible to investigate further on their association.

\section{Software and third party data repository citations} \label{sec:cite}
The \textit{Fermi}-LAT $\gamma$-ray data analysis was done with `Fermipy' \citep{2017ICRC...35..824W}. Swift X-ray, Ultraviolet \& Optical data have been analyzed with `HEASoft' \citep{2014ascl.soft08004N}. To analyze XMM-Newton data, we have used Science Analysis System (SAS; \cite{2004ASPC..314..759G}).

\begin{acknowledgments}
We thank the referee for helpful comments to improve the paper.
 S.K.M. thanks T. Ghosh, Hemanth M., A. D. Sarkar for useful discussions. 
This research has made use of data from the MOJAVE database that is maintained by the MOJAVE team \citep{2018ApJS..234...12L}.
\end{acknowledgments}

\facilities{Swift(XRT and UVOT), XMM-Newton, \textit{Fermi}-LAT, MAGIC.}

\software{Fermipy (\url{https://fermipy.readthedocs.io/en/latest/}; \cite{2017ICRC...35..824W}), HEASoft (\url{https://heasarc.gsfc.nasa.gov/docs/software/lheasoft/}; \cite{2014ascl.soft08004N}), SAS (\url{https://www.cosmos.esa.int/web/xmm-newton/sas-threads}; \cite{2004ASPC..314..759G}), GAMERA (\url{http://libgamera.github.io/GAMERA/docs/main_page.html}; \cite{Hahn:2015hhw}), CRPropa 3 (\url{https://crpropa.github.io/CRPropa3/}; \cite{AlvesBatista:2016vpy, AlvesBatista:2022vem})}.


\bibliography{sample631}{}
\bibliographystyle{aasjournal}



\begin{table*}
\caption{Results of multi-wavelength SED modeling shown in the Fig. ~\ref{fig:Pure_LP_MWSED}, ~\ref{fig:Lp_UHECR_MWSED} and ~\ref{fig:LP_PP_MWSED}}
\centering
\begin{tabular}{cccc rrrrr}   
\hline\hline
 Parameters & & Values &  &\\
\hline
& Pure-leptonic  & Leptonic+ hadronic (UHECR) & Leptonic+ hadronic ($pp$) \\
& model & model & model \\
\hline
 Spectral index of injected electron spectrum ($\alpha$) &  2.2 & 2.2& 2.2&\\
  Curvature index of injected electron spectrum  ($\beta$) & 0.06& 0.06 & 0.10 &\\
 Magnetic field in emission region (B) &  0.10 G & 0.10 G & 0.10 G &\\
 Size of the emission region (R) &  8.0$\times10^{15}$ cm & 8.0$\times10^{15}$ cm& 1.8$\times10^{16}$ cm& \\
 Doppler factor ($\delta_D$) &  20 & 20& 20& \\
 Min. Lorentz factor ($\gamma_{\rm min}$)  & 1.0$\times10^2$ & 1.0$\times10^2$& 2.5$\times10^2$& \\
 Max. Lorentz factor ($\gamma_{\rm max}$)  & 9.0$\times10^7$ & 9.0$\times10^7$ & 9.0$\times10^7$& \\
 Spectral index of relativistic proton spectrum ($\alpha_p$) &-- & 2.2 & 2.2\\
  Min. energy of relativistic protons (E$_{\rm p,min} $) & -- & 0.1 EeV & 10 GeV\\
 Max. energy of relativistic protons (E$_{\rm p,max} $) & -- & 100 EeV & $10^4$ GeV\\
 Jet power of relativistic leptons (P$_e$) &  2.6$\times10^{43}$ erg/s & 2.6$\times10^{43}$ erg/s & 2.2$\times10^{43}$ erg/s & \\
Jet power of magnetic field  (P$_B$) &  9.6$\times10^{41}$ erg/s & 9.6$\times10^{41}$ erg/s & 4.9$\times10^{42}$ erg/s& \\
 Jet power of relativistic protons (P$_p$) & -- & 1.9$\times10^{42}$ erg/s & 9.8$\times10^{44}$ erg/s &\\
 Kinematic jet power (P$^k_{\rm tot}$) &  2.7$\times10^{43}$ erg/s & 2.9 $\times10^{43}$ erg/s & $1.0\times10^{45}$ erg/s& \\
\hline

\end{tabular}
\label{tab:GAMERA_Fitting_param}
\end{table*}

\end{document}